\documentclass[preprint,12pt]{elsarticle}
\usepackage{amsmath}
\usepackage{amssymb}

\usepackage{mathtools}
\usepackage[hidelinks]{hyperref}
\hypersetup{pdftitle = {Surface roughness in finite element meshes}}
\usepackage{cleveref}

\graphicspath{ {./figures/} }

\newcommand{\expval}[1]{\langle #1 \rangle} %%% expectation value
\newcommand{\transpose}{^{\mathrm{T}}}

\usepackage[dvipsnames]{xcolor}

\journal{Journal of Computational Physics}

\begin{document}

\begin{frontmatter}

\title{Surface roughness in finite element meshes}

\author[hu,mbi]{Fabian Loth\corref{cor1}}
\ead{loth@physik.hu-berlin.de}

\author[hu]{Thomas Kiel}
\author[hu,mbi]{Kurt Busch}
\author[hu]{Philip Tr{\o}st Kristensen}
\cortext[cor1]{Corresponding author.}

\address[hu]{Humboldt-Universit{\"a}t zu Berlin, 
Institut f{\"u}r Physik, AG Theoretische Optik \& Photonik, 
Newtonstra{\ss}e 15, 12489 Berlin, Germany}

\address[mbi]{Max-Born-Institut, Max-Born-Stra{\ss}e 2A, 12489 Berlin, Germany}

\begin{abstract}
We present a practical approach for constructing meshes of general rough surfaces with given autocorrelation functions based on the unstructured meshes of nominally smooth surfaces.
The approach builds on a well-known method to construct correlated random numbers from white noise using a decomposition of the autocorrelation matrix. 
We discuss important details arising in practical applications to the physicalmodeling of surface roughness and provide a software implementation to enable use of the approach with a broad range of numerical methods in various fields of science and engineering.
\end{abstract}


\end{frontmatter}

%% \linenumbers

%% main text
\section{Introduction}
Surface roughness plays an important role in physical and chemical phenomena and is, therefore, of great importance in science and engineering~\cite{gong2018}.
Indeed, in the relatively broad area of nanophotonics alone with which we have first-hand modeling experience, it is known that the optical response of any structure can, in general, be significantly changed by surface roughness. 
For instance, this plays a role in light scattering by small particles~\cite{li2004}, influences the optical properties of plasmonic nanostructures~\cite{truegler2011,truegler2014,lu2018}, leads to detrimental losses in photonic crystal waveguides~\cite{johnson2005}, modifies the performance of hyperbolic metamaterials~\cite{kozik2014} and affects the Casimir force~\cite{vanZwol2011}.
Moreover, it is by now well recognized, that irregularities and protrusions on general metallic surfaces can lead to the formation of hot-spots with enormous optical field enhancements which, in turn, enables surface-enhanced Raman spectroscopy~\cite{Fleishmann_1974} with applications ranging from biophysics~\cite{Kneipp_2002} to the detection of trace materials such as explosives~\cite{Hakonen_2015}.
Furthermore, surface roughness influences the photon extraction of light emitting diodes \cite{Fujii2004} and the quantum efficiency of solar cells~\cite{Krc2003}.
Apart from applications in nanophotonics, it is known that scattering of phonons at rough surfaces of materials can reduce the thermal conductivity~\cite{Santamore2001}, just as surface roughness has a decisive influence on qualitative as well as quantitative details of fluid dynamics, in particular in the vast research field of micro- and nanofluidics~\cite{Taylor2005}.
Lately, it was demonstrated that surface roughness can increase the catalytic activity of materials~\cite{Kim2019} and theoretical studies show that surface roughness affects the electron confinement in semiconductor quantum dots~\cite{Macdo2012}.
As a last example of our non-exhaustive list, we note that surface roughness plays a key role in the tribology of micro- and nanoelectromechanical systems~\cite{Bhushan2005}.

Practical characterizations and measurements of surface roughness at the micro- and nanoscale can be performed with techniques such as atomic force microscopy, confocal laser scanning microscopy, scanning interferometry and scatterometry~\cite{gong2018}.
This data can then be used in the physical modeling of experiments, for example, to asses if the measured or expected levels of surface roughness can explain the deviations from the theoretical predictions which are usually based on idealized smooth geometries.
Similarly, the characteristics can be used in the design of devices to model if the expected or measured surface roughness is likely to degrade the performance of the device.
In practice, such modeling is almost exclusively done by numerical means, and therefore it is of some interest to explore and discuss the practical generation of finite element meshes for general rough surfaces with given autocorrelation functions based on unstructured meshes of nominally smooth surfaces.

Surface roughness is the deviation in normal direction of a real surface from its nominal form and is typically characterized by the root mean squared (rms) roughness $R_{\mathrm{q}}$ in the normal direction of the nominal form and the correlation length $l$ in the lateral (tangential) direction of the nominal form.
Actually, a mesh with given rms roughness can be easily achieved by simply moving points on the mesh of the nominal surface in the normal direction by an amount set by uncorrelated random numbers.
However, such a naive approach produces roughness with a vanishing correlation length, and this is far from a good model for the roughness of typical real surfaces.
In order to introduce a finite correlation length, one option is to perform a convolution of the uncorrelated random numbers with a filter function of the desired form~\cite{garcia1984}. 
This method, which is known as the spectral method~\cite{Warnick2001}, is suitable for regular grids of rectangular surfaces and produces periodic roughness.
For instance, it has been used for the computation of the absorption of light by rough metal surfaces~\cite{Bergstrm2008}.
As an alternative to the spectral method, we discuss in this article how one can instead work directly with the unstructured meshes of nominal structures.
Such unstructured meshes are typical for finite element techniques.

\begin{figure}
\includegraphics[width=\linewidth]{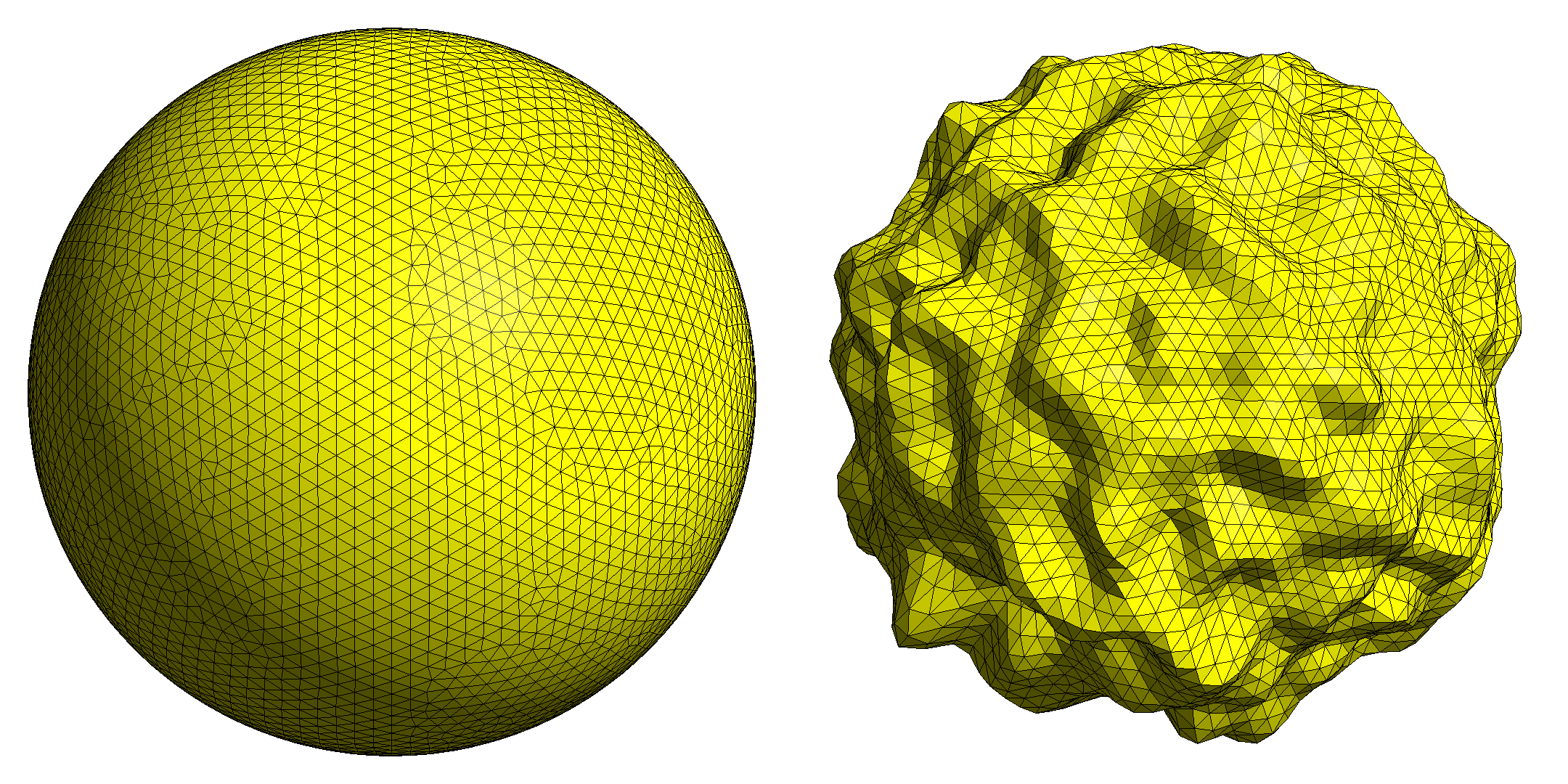}
\caption{Unstructured mesh of a sphere with radius $r$ generated with Gmsh~\cite{Geuzaine2009} and a rough sphere with rms roughness $R_{\mathrm{q}}/r= 0.05$, correlation length $l/r = 0.15$ and element size $h/r=0.05$, which is called \emph{characteristic length} in Gmsh.}
\label{fig:rough_sphere}
\end{figure}

Specifically, we create correlated numbers by using a well-known method based on decomposition of the autocorrelation matrix~\cite{Kaiser1962,Gallager2013}.
This method is useful for a relatively broad range of three-dimensional geometries, including cases which cannot be constructed from a rectangle without distorting the distances between the points on the surface.
In particular, we apply it to generate rough spheres, as illustrated in~\cref{fig:rough_sphere}, which shows the nominal mesh of a sphere generated with Gmsh~\cite{Geuzaine2009} along with a single realization of a corresponding rough mesh.

This article is organized as follows.
In \cref{sec:method}, we present the method for constructing meshes of general rough surfaces, and \cref{sec:details} contains details of the method in order to use it in practice.
Finally, \cref{sec:conclusion} holds the conclusions. 

\section{Method}
\label{sec:method}
Given a nominally smooth surface, we describe the local deviation in the normal direction as a stationary Gaussian stochastic process $D=D(t)$.
Each realization of the stochastic process yields an ordinary function $d(t)$, where $t$ denotes the tangential coordinate along the surface.
In the discrete case considered in this work, each $t_i$ denotes the position of a vertex in the nominal mesh and $d(t_i) = d_i$ denotes the corresponding shift in normal direction.
The mean of the Gaussian process $\langle D \rangle=0$ vanishes, as it corresponds to the nominal surface.
For simplicity we can set the variance to one and multiply the deviation with the desired root mean squared roughness $R_{\mathrm{q}}$ at the end of the calculation.
As a consequence, the autocovariance, the autocorrelation and the second moment are equivalent for this stochastic process.

\begin{figure}
\includegraphics[width=\textwidth]{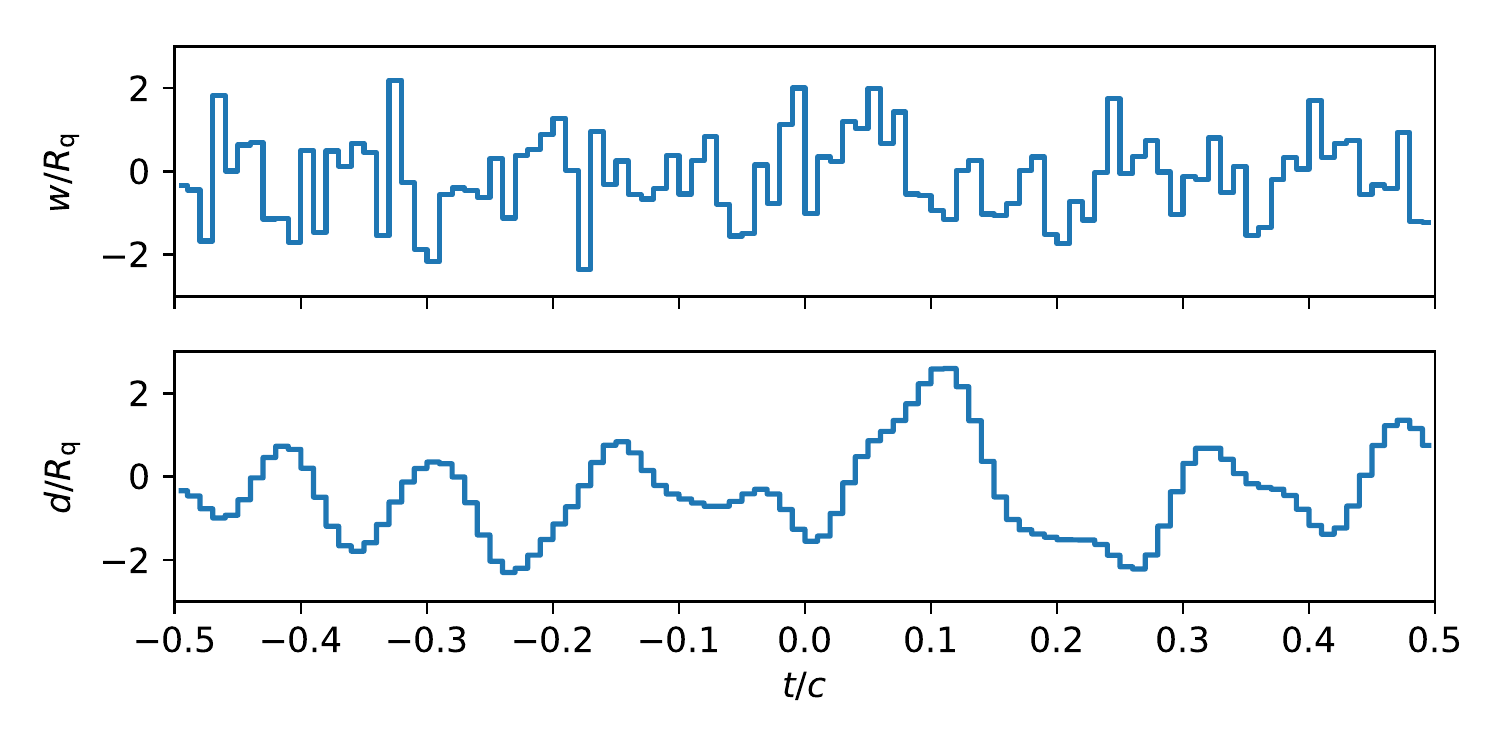}
\vspace{-1cm}
\caption{Discrete Gaussian white noise $w$ with $n=100$ points generated with a random number generator (top) and the height deviation $d$ of a one-dimensional rough surface created using a Gaussian autocorrelation with correlation length $l/c = 0.03$ (bottom).
The tangential coordinate $t$ and the amplitudes $w$ and $d$ are normalized to the total length $c$ and the root mean squared roughness $R_{\mathrm{q}}$, respectively.}
\label{fig:rough_surface_1d}
\end{figure}

We start from the simplest stationary Gaussian stochastic process, which is known as Gaussian white noise $W$ and which is determined by its autocorrelation matrix
\begin{equation}
\expval{WW\transpose} =I \, , \quad I_{ij} = \delta_{ij} \, .
\end{equation}
Each realization $w$ is a vector of normally distributed uncorrelated random numbers, which, in practice, can be conveniently generated with a pseudo-random number generator, such as the Mersenne Twister~\cite{Matsumoto1998} used in this work~\cite{Oliphant2017}. 
One example realization in one dimension, which uses $w$ directly as height deviation, is depicted in the top panel of~\cref{fig:rough_surface_1d}.
It clearly shows that Gaussian white noise is a rather bad model for most numerical models of roughness.
Not only is the lack of a finite correlation length in lateral direction in principle inconsistent with reality, but the discrete realization in practice enforces a correlation length equal to the mesh size.
This latter property fundamentally changes the convergence characteristics of any numerical scheme relying on refinement of the calculation mesh to increase accuracy.

\begin{figure}
\includegraphics[width=\linewidth]{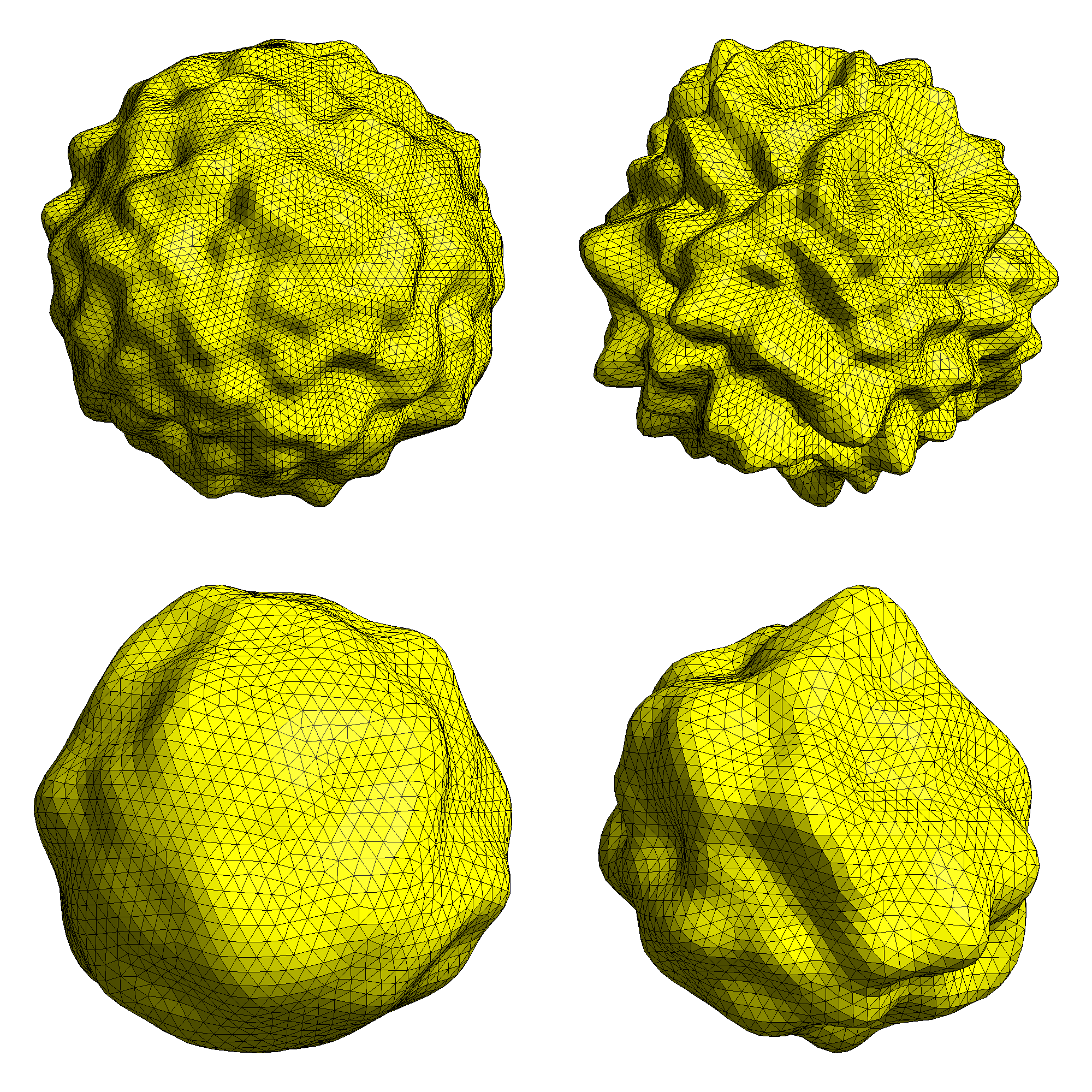}
\caption{Meshes of spheres with nominal radius $r$ and different roughness.
The rms roughness is doubled from $R_{\mathrm{q}}/r = 0.04$ (left) to $R_{\mathrm{q}}/r = 0.08$ (right) and the correlation length is doubled from $l/r = 0.1$ (top) to $l/r = 0.2$ (bottom).
The element size is $h = l / 2.5$.}
\label{fig:four_spheres}
\end{figure}

Even if Gaussian white noise is a bad model for most, if not all, realistic cases of roughness, it is a wonderful fact, that we can use it to construct any other Gaussian random vector~\cite{Gallager2013}.
For the examples in this work, we assume that the height deviation $D$ has a Gaussian autocorrelation matrix of the form
\begin{equation}
    \label{eq:gaussian_autocorrelation}
    \expval{DD\transpose} = R \, , \quad
    R_{ij} = \exp \left(-\frac{|t_i - t_j|^2}{2 l^2} \right) \, ,
\end{equation}
but we note that other forms of correlation, e.g.\ exponential correlation~\cite{Ogilvy1989}, can be treated immediately by simply inserting the desired functional form for $R_{ij}$.
To construct correlated random numbers from Gaussian white noise, we seek a matrix $L$ such that $D=LW$.
Substitution into the autocorrelation yields:
\begin{equation}
  R = \expval{DD\transpose}
    = \expval{(LW)(LW)\transpose}
    = \expval{LWW\transpose L\transpose}
    = L \expval{WW\transpose} L\transpose
    = LL\transpose \, ,
\end{equation}
from which it follows, that any decomposition of the desired autocorrelation matrix into a matrix $L$ and its transpose $L\transpose$ will suffice.
As stated in the introduction, this is a well-known approach for generating correlated numbers in general~\cite{Kaiser1962,Gallager2013}, and we apply it here to generate surface roughness in finite element meshes.
The matrix $L$ can be obtained by the Cholesky decomposition $R=LL\transpose$, where $R$ and $L$ are, respectively, symmetric positive-definite and lower triangular matrices~\cite{higham2009}.
We note that $R$ is symmetric by construction and for now we assume that it is also positive-definite; this assumption is further discussed in~\cref{sec:eigenvalues}.
The method can be directly applied to unstructured meshes of general surfaces by defining $|t_i-t_j|$ to be the distance between vertices $i$ and $j$.
It proceeds in eight distinct steps as follows:
\begin{enumerate}
  \item Generate a mesh of the nominal surface using a mesh generator.
  \item Compute the pairwise distances between all vertices.
  \item Construct the desired autocorrelation matrix $R$.
  \item Compute the Cholesky decomposition $R=LL\transpose$.
  \item Generate Gaussian white noise $w$ using a random number generator.
  \item Calculate the height deviation $d=Lw$ with the desired autocorrelation.
  \item Multiply $d$ by the desired rms roughness to obtain $\tilde{d}=R_{\mathrm{q}} d$.
  \item Shift all vertices in normal direction according to the elements of $\tilde{d}$.
\end{enumerate}
This approach, which is implemented in a collection of Python scripts and published together with this article \cite{roughmesh}, was used to generate the rough sphere in~\cref{fig:rough_sphere}.
The effect of different roughness parameters is illustrated in ~\cref{fig:four_spheres}.
It shows, that the root mean squared roughness $R_{\mathrm{q}}$ controls the amplitude of the roughness while the correlation length $l$ is proportional to the typical distance between bumps on the surface. 
Moreover, it illustrates that one typically needs a finer mesh for shorter correlation lengths.
\section{Details}
\label{sec:details}
Even if the principles of the suggested approach are relatively simple, its practical implementation calls for the discussion of a number of details.
\subsection{Sampling}
Describing surface roughness as a discrete stochastic process is an approximation to the actual continuous rough surface.
We examine the quality of this approximation by means of the sampling theorem~\cite{shannon1949}, which states that a function which is bandwidth-limited to frequencies smaller than $f_{\mathrm{max}}$ can be reconstructed exactly from its equidistantly sampled values if the sampling frequency is larger than $2 f_{\mathrm{max}}$.

By means of the Wiener-Khinchin theorem, the spectral density $S(f)$ of a stationary stochastic process is given by the Fourier transform of the autocorrelation function~\cite{Wiener1930,Khintchine1934}.
Thus, for a Gaussian autocorrelation the spectral density $S(f)$ is Gaussian and not strictly bandwidth-limited.
Nevertheless, we may define a frequency $f_{\mathrm{max}}$, such that $S(f)$ is negligibly small for $f>f_{\mathrm{max}}$.
Then, for a sampling frequency of $2 f_{\mathrm{max}}$, the surface in principle can be reconstructed with negligible aliasing using the Whittaker-Shannon interpolation formula~\cite{shannon1949}.
In this work, however, we are interested in the construction of meshes for use with numerical methods such as finite element methods, wherefore we are typically restricted to linear interpolations of the calculation domain into triangular or tetrahedral elements.
In this case the discretization in general needs to be finer to represent high frequency fluctuations correctly, cf.\ \cref{fig:four_spheres}.
Since a finer mesh increases the size of the numerical problem, one needs to find a compromise between accuracy and computation time depending on the problem at hand.
Even if this is a general characteristic of the numerical calculations, the introduction of roughness will typically lead to an inherent uncertainty in the final results and thereby add a natural bound on the calculation accuracy beyond which it does not make sense to refine the calculation mesh further.
Indeed, if one finds that the introduction of roughness leads to a $10$ percent variation in the figure of merit, one can safely perform calculations with an estimated error of one percent.
In the examples in \cref{fig:rough_sphere,fig:four_spheres}, the element size is chosen as $h=l/2.5$, where $l$ is the correlation length.

\subsection{Eigenvalues of the autocorrelation matrix}
\label{sec:eigenvalues}
\begin{figure}
\includegraphics[width=\textwidth]{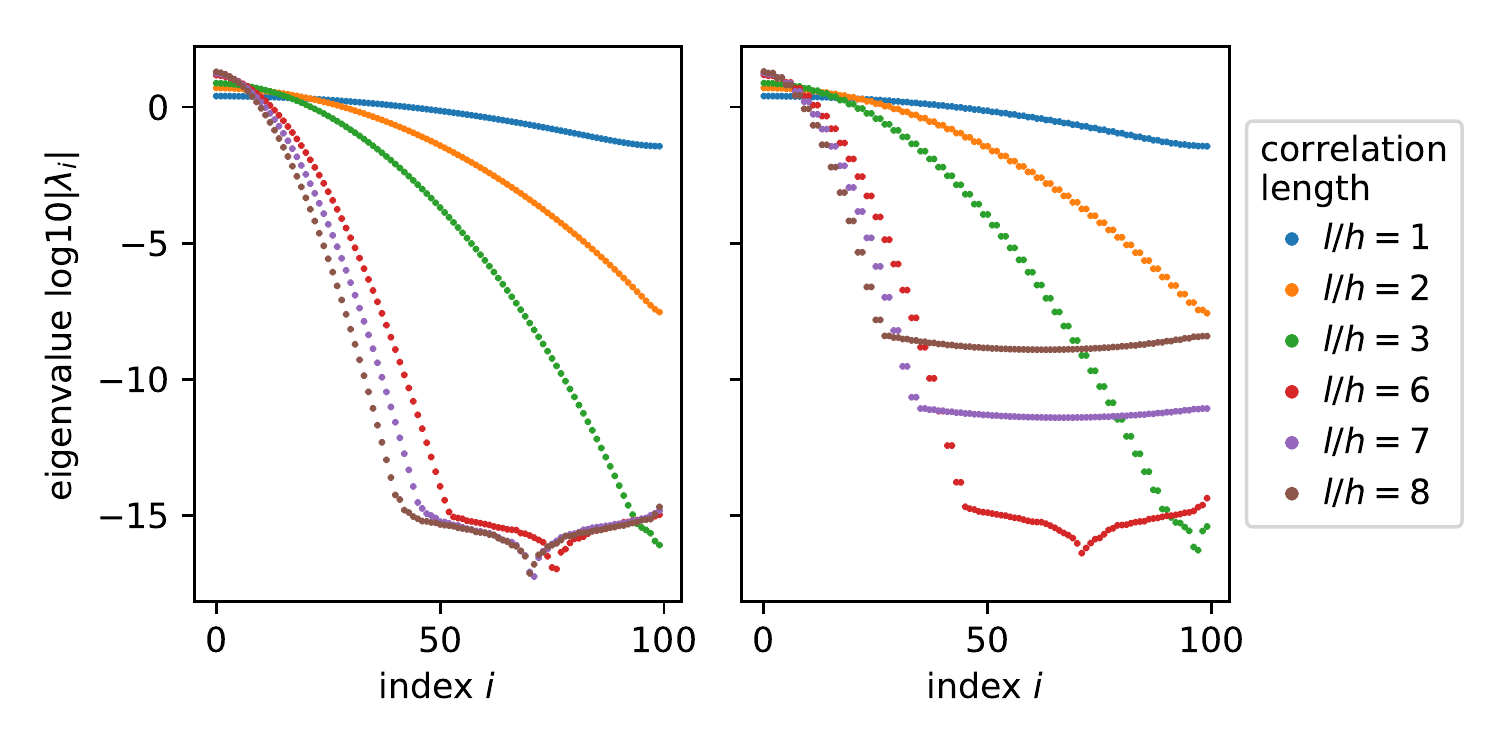}
\caption{Logarithm of the computed eigenvalues $\lambda_i$ for the Gaussian  autocorrelation matrix with different correlation lengths $l$ in descending order for an open (left) and a periodically closed (right) one-dimensional surface of length $c$ and sampled with fixed element size $h/c = 0.01$.
For correlation lengths $l/h \gtrsim 3$, points on the right side of the minimum correspond to negative eigenvalues.}
\label{fig:eigval_1d}
\end{figure}

The Cholesky decomposition of a symmetric matrix exists if and only if the matrix is positive-definite meaning that all eigenvalues are strictly positive.
Thus, if the numerical Cholesky decomposition of the autocorrelation matrix succeeds, then the matrix is positive-definite.
Furthermore, Schoenberg proved that the function $\exp(-|x|^p)$ with $x\in \mathbb{R} $ is positive definite if $0<p\leq2$ and not positive definite if $p>2$~\cite{Schoenberg1938}.
We have found, however, that under certain conditions the numerical Cholesky decomposition fails even for $p=2$, which corresponds to the Gaussian autocorrelation given by~\cref{eq:gaussian_autocorrelation}.
We examine this phenomenon by computing the eigenvalues of the Gaussian autocorrelation matrix with different correlation lengths $l$ for a one-dimensional surface of length $c$ sampled with fixed element size $h/c=0.01$, see~\cref{fig:eigval_1d}, left.
First, we notice that the symmetric eigenvalue problem is well-conditioned~\cite{Bauer1960} even if the matrix itself is ill-conditioned as it is the case for the autocorrelation matrix.
Because of Szeg{\"o}'s theorem~\cite{Moon2000}, we expect the eigenvalues of the autocorrelation matrix to have the same distribution as the spectral density.
Indeed, the eigenvalue spectrum has a normal distribution which shows a parabola in the logarithmic plot in \cref{fig:eigval_1d}.
As the eigenvalues reach the machine precision $\epsilon \approx 10^{-16}$, however, all smaller eigenvalues are distributed around zero with an absolute value on the order of $\epsilon$.
This happens for $l/h \gtrsim 3 $, and in these cases, the Cholesky decomposition fails.
Instead, we compute the eigendecomposition
\begin{equation}
  R = Q \, \Lambda \, Q\transpose \, ,
\end{equation}
where $Q$ is orthogonal and $\Lambda$ is diagonal and contains the eigenvalues.
We take the positive part $\Lambda^+$ by setting all negative eigenvalues to zero, 
and use $Q\sqrt{\Lambda^+}$ instead of the Cholesky factor $L$.

\begin{figure}
\includegraphics[width=\textwidth]{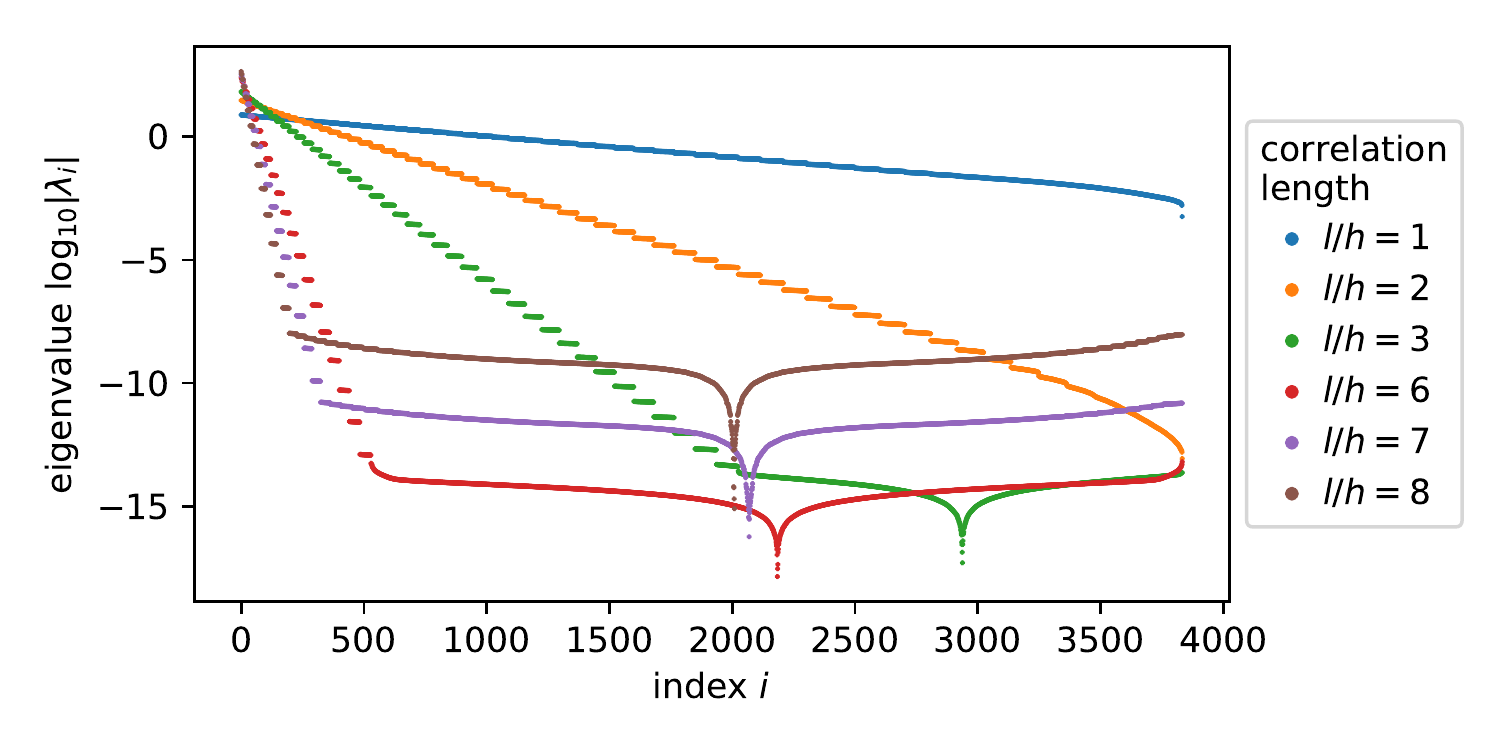}
\caption{Logarithm of the computed eigenvalues $\lambda_i$ of the Gaussian autocorrelation matrix with different correlation length $l$ in descending order for a spherical mesh with circumference $c$ and element size $h/c = 0.01$.
For correlation lengths $l/h \gtrsim 3$, points on the right side of the minimum correspond to negative eigenvalues.}
\label{fig:eigval_sphere}
\end{figure}

An additional complication may arise in the case of periodic boundary conditions, as seen in the right panel of~\cref{fig:eigval_1d}.
In this case, the theorem by Schoenberg does not apply, thus, it is not guaranteed that the autocorrelation matrix is positive-definite.
The relevant quantity for the following discussion is the ratio of the correlation length $l$ and the circumference $c$.
For the cases $l/c \gtrsim 0.07$ (which corresponds in~\cref{fig:eigval_1d} to $l/h \gtrsim 7$), the Gaussian eigenvalue spectrum is cut off at a value $\lambda_{\mathrm{cut}} \gg \epsilon$, after which the eigenvalues are distributed around zero with absolute values of the order of $\lambda_{\mathrm{cut}}$.
This effect is visible in the spectra as a number of distinct floors, at which the magnitude of the eigenvalues are approximately constant.
It turns out that this value is of the same order as the minimal correlation, i.e.\ the correlation between most distant points in the mesh.
To avoid this problem, we need to chose the correlation length such that the minimal correlation is smaller than the desired precision.
Fortunately, this is inherently unproblematic, as a strong correlation of all points would essentially constitute a different geometry than that of a rough surface.
Clearly, these considerations hold also for two-dimensional surfaces, e.g.\ a sphere (\cref{fig:eigval_sphere}).
As a side note, we see that the eigenvalue spectrum has steps of almost the same values.
The width of these steps increases such that the spectrum is linear in the logarithmic plot.
This corresponds to the fact that for each vertex, the distance to surrounding vertices is roughly an integer multiple of the element size and the number of vertices with distance $d$ increases as $d$ increases itself.

\subsection{Distance matrix}
\label{sec:distance}
In order to construct the autocorrelation matrix one needs to compute the pairwise distances between all vertices of the nominal surface mesh.
On curved surfaces, it is arguably reasonable to interpret this as the geodesic distance, i.e.\ the shortest path along the surface.
Analytic expressions for the geodesic distance, however, are only available for simple geometries, such as spheres. 
For other geometries, we use the Euclidean distance as an approximation.
This approximation is expected to be best in the limit of short correlation lengths relative to the curvature of the surface.
In particular, sharp edges and corners should be avoided.
To handle such geometries, one may have to use an additional step to first round the corners, before applying the roughness.
We note, that the geodesic distance between points on an arbitrary triangular mesh may be computed using an algorithm by Mitchell, Mount and Papadimitrou~\cite{mitchell1987} and variants of it~\cite{surazhsky2005}.
However, it is much more costly than the simple Euclidean distance and thus it was not used in our present work.

\section{Conclusion}
\label{sec:conclusion}
We have presented a practical approach for constructing finite element meshes of general rough surfaces with a desired autocorrelation based on a decomposition of the autocorrelation matrix.
In addition, we have discussed important details of the method and especially explained and solved the problem of negative eigenvalues of the autocorrelation matrix.
The approach can be directly used with the code provided together with this article \cite{roughmesh}.

\section*{Acknowledgment}
This work was supported by the German Federal Ministry of Education and Research through the funding program Photonics Research Germany (Project 13N14149).
F.L. and K.B. further acknowledge the support by Deutsche Forschungsgemeinschaft (DFG) within the priority program SPP 1839 "Tailored Disorder" (project BU 1107/10-2).

\bibliographystyle{unsrt}
\bibliography{bibliography}

\begin{thebibliography}{10}

\bibitem{gong2018}
Y.~Gong, J.~Xu, and R.C. Buchanan.
\newblock Surface roughness: A review of its measurement at micro-/nano-scale.
\newblock {\em Physical Sciences Reviews}, 3(1), 2018.

\bibitem{li2004}
C.~Li, G.W. Kattawar, and P.~Yang.
\newblock Effects of surface roughness on light scattering by small particles.
\newblock {\em J. Quant. Spectrosc. Radiat. Transfer}, 89(1-4):123--131, 2004.

\bibitem{truegler2011}
A.~Tr{\"u}gler, J.-C. Tinguely, J.R. Krenn, A.~Hohenau, and U.~Hohenester.
\newblock Influence of surface roughness on the optical properties of plasmonic
  nanoparticles.
\newblock {\em Phys. Rev. B}, 83(8):081412, 2011.

\bibitem{truegler2014}
A.~Tr\"ugler, J.-C. Tinguely, G.~Jakopic, U.~Hohenester, J.R. Krenn, and
  A.~Hohenau.
\newblock Near-field and sers enhancement from rough plasmonic nanoparticles.
\newblock {\em Phys. Rev. B}, 89(16):165409, 2014.

\bibitem{lu2018}
Y.-W. Lu, L.-Y. Li, and J.-F. Liu.
\newblock Influence of surface roughness on strong light-matter interaction of
  a quantum emitter-metallic nanoparticle system.
\newblock {\em Sci. Rep.}, 8(1), 2018.

\bibitem{johnson2005}
S.G. Johnson, M.L. Povinelli, M.~Soljacic, A.~Karalis, S.~Jacobs, and J.D.
  Joannopoulos.
\newblock Roughness losses and volume-current methods in photonic-crystal
  waveguides.
\newblock {\em Applied Physics B}, 81(2-3):283--293, 2005.

\bibitem{kozik2014}
S.~Kozik, M.A. Binhussain, A.~Smirnov, N.~Khilo, and V.~Agabekov.
\newblock Investigation of surface roughness influence on hyperbolic
  metamaterial performance.
\newblock {\em Advanced Electromagnetics}, 3(2), 2014.

\bibitem{vanZwol2011}
P.~J. van Zwol, V.~B. Svetovoy, and G.~Palasantzas.
\newblock Characterization of optical properties and surface roughness
  profiles: The casimir force between real materials.
\newblock In {\em Casimir Physics}, pages 311--343. Springer Berlin Heidelberg,
  2011.

\bibitem{Fleishmann_1974}
M.~Fleischmann, P.J. Hendra, and A.J. McQuillan.
\newblock Raman spectra of pyridine adsorbed at a silver electrode.
\newblock {\em Chemical Physics Letters}, 26(2):163 -- 166, 1974.

\bibitem{Kneipp_2002}
Katrin Kneipp, Harald Kneipp, Irving Itzkan, Ramachandra~R Dasari, and
  Michael~S Feld.
\newblock Surface-enhanced raman scattering and biophysics.
\newblock {\em Journal of Physics: Condensed Matter}, 14(18):R597--R624, 2002.

\bibitem{Hakonen_2015}
Aron Hakonen, Per~Ola Andersson, Michael~Stenb{\ae}k Schmidt, Tomas
  Rindzevicius, and Mikael K{\"a}ll.
\newblock Explosive and chemical threat detection by surface-enhanced raman
  scattering: A review.
\newblock {\em Analytica Chimica Acta}, 893:1 -- 13, 2015.

\bibitem{Fujii2004}
T.~Fujii, Y.~Gao, R.~Sharma, E.~L. Hu, S.~P. DenBaars, and S.~Nakamura.
\newblock Increase in the extraction efficiency of {GaN}-based light-emitting
  diodes via surface roughening.
\newblock {\em Applied Physics Letters}, 84(6):855--857, February 2004.

\bibitem{Krc2003}
J.~Kr{\v{c}}, M.~Zeman, O.~Kluth, F.~Smole, and M.~Topi{\v{c}}.
\newblock Effect of surface roughness of {ZnO}:al films on light scattering in
  hydrogenated amorphous silicon solar cells.
\newblock {\em Thin Solid Films}, 426(1-2):296--304, February 2003.

\bibitem{Santamore2001}
D.~H. Santamore and M.~C. Cross.
\newblock Effect of surface roughness on the universal thermal conductance.
\newblock {\em Physical Review B}, 63(18), April 2001.

\bibitem{Taylor2005}
James~B. Taylor, Andres~L. Carrano, and Satish~G. Kandlikar.
\newblock Characterization of the effect of surface roughness and texture on
  fluid flow: Past, present, and future (keynote).
\newblock In {\em {ASME} 3rd International Conference on Microchannels and
  Minichannels, Parts A and B}. {ASME}, 2005.

\bibitem{Kim2019}
Ikhyun Kim, Gisu Park, and Jae~Jeong Na.
\newblock Experimental study of surface roughness effect on oxygen catalytic
  recombination.
\newblock {\em International Journal of Heat and Mass Transfer}, 138:916--922,
  August 2019.

\bibitem{Macdo2012}
R.~Mac{\^{e}}do, M.~S. Sena, J.~Costa e~Silva, A.~Chaves, and J.~A.~P.
  da~Costa.
\newblock The role of surface roughness on the electron confinement in
  semiconductor quantum dots.
\newblock In {\em Latin America Optics and Photonics Conference}. {OSA}, 2012.

\bibitem{Bhushan2005}
Bharat Bhushan.
\newblock Micro/nanotribology of {MEMS}/{NEMS} materials and devices.
\newblock In {\em Nanotribology and Nanomechanics}, pages 1031--1089.
  Springer-Verlag, 2005.

\bibitem{garcia1984}
N.~Garcia and E.~Stoll.
\newblock Monte carlo calculation for electromagnetic-wave scattering from
  random rough surfaces.
\newblock {\em Phys. Rev. Lett.}, 52(20):1798--1801, 1984.

\bibitem{Warnick2001}
Karl~F Warnick and Weng~Cho Chew.
\newblock Numerical simulation methods for rough surface scattering.
\newblock {\em Waves in Random Media}, 11(1):R1--R30, January 2001.

\bibitem{Bergstrm2008}
D.~Bergstr\"{o}m, J.~Powell, and A.~F.H. Kaplan.
\newblock The absorption of light by rough metal surfaces{\textemdash}a
  three-dimensional ray-tracing analysis.
\newblock {\em Journal of Applied Physics}, 103(10):103515, May 2008.

\bibitem{Kaiser1962}
H.F. Kaiser and K.~Dickman.
\newblock Sample and population score matrices and sample correlation matrices
  from an arbitrary population correlation matrix.
\newblock {\em Psychometrika}, 27(2):179--182, 1962.

\bibitem{Gallager2013}
Robert~G. Gallager.
\newblock {\em Stochastic Processes}.
\newblock Cambridge University Press, December 2013.

\bibitem{Geuzaine2009}
C.~Geuzaine and J.-F. Remacle.
\newblock Gmsh: A 3-d finite element mesh generator with built-in pre- and
  post-processing facilities.
\newblock {\em International Journal for Numerical Methods in Engineering},
  79(11):1309--1331, 2009.

\bibitem{Matsumoto1998}
Makoto Matsumoto and Takuji Nishimura.
\newblock Mersenne twister: A 623-dimensionally equidistributed uniform
  pseudo-random number generator.
\newblock {\em ACM Transactions on Modeling and Computer Simulation},
  8(1):3--30, jan 1998.

\bibitem{Oliphant2017}
T.~E. {Oliphant}.
\newblock Python for scientific computing.
\newblock {\em Computing in Science Engineering}, 9(3):10--20, May 2007.

\bibitem{Ogilvy1989}
J.A. Ogilvy and J.R. Foster.
\newblock Rough surfaces: gaussian or exponential statistics?
\newblock {\em Journal of Physics D: Applied Physics}, 22(9):1243--1251, 1989.

\bibitem{higham2009}
N.~J. Higham.
\newblock Cholesky factorization.
\newblock {\em WIREs Comp. Stat.}, 1:251--254, 2009.

\bibitem{roughmesh}
F.~Loth.
\newblock roughmesh.
\newblock https://github.com/fabian-loth/roughmesh, 2020.

\bibitem{shannon1949}
C.E. Shannon.
\newblock Communication in the presence of noise.
\newblock {\em Proc. IRE}, 37(1):10--21, 1949.

\bibitem{Wiener1930}
Norbert Wiener.
\newblock Generalized harmonic analysis.
\newblock {\em Acta Mathematica}, 55(0):117--258, 1930.

\bibitem{Khintchine1934}
A.~Khintchine.
\newblock Korrelationstheorie der station{\"a}ren stochastischen prozesse.
\newblock {\em Mathematische Annalen}, 109(1):604--615, December 1934.

\bibitem{Schoenberg1938}
I.J. Schoenberg.
\newblock Metric spaces and positive definite functions.
\newblock {\em Transactions of the American Mathematical Society},
  44(3):522--522, March 1938.

\bibitem{Bauer1960}
F.~L. Bauer and C.~T. Fike.
\newblock Norms and exclusion theorems.
\newblock {\em Numerische Mathematik}, 2(1):137--141, December 1960.

\bibitem{Moon2000}
T.K. Moon and W.C. Sterling.
\newblock {\em Mathematical Methods and Algorithms for Signal Processing}.
\newblock Prentice Hall, 2000.

\bibitem{mitchell1987}
J.S.B. Mitchell, D.M. Mount, and C.H. Papadimitriou.
\newblock The discrete geodesic problem.
\newblock {\em SIAM J. of Computing}, 16(4):647--668, 1987.

\bibitem{surazhsky2005}
V.~Surazhsky, T.~Surazhsky, D.~Kirsanov, S.J. Gortler, and H.~Hoppe.
\newblock Fast exact and approximate geodesics on meshes.
\newblock {\em ACM Trans. Graph.}, 24(3):553--560, 2005.

\end{thebibliography}

%\printbibliography

\end{document}